\documentclass[11pt]{article}
\usepackage[a4paper,margin=1in]{geometry}
\usepackage[T1]{fontenc}
\usepackage[utf8]{inputenc}
\usepackage{amsmath,amssymb}
\usepackage{graphicx}
\usepackage{xcolor}
\usepackage{booktabs}
\usepackage{array}
\usepackage{url}
\usepackage{tabularx}
\usepackage[authoryear,round]{natbib}
\usepackage{authblk}
\usepackage[hidelinks]{hyperref}

\newenvironment{keywords}
  {\par\small\noindent\textit{Keywords: }\ignorespaces}
  {\par}
\title{{\color{black}Öpik-type collision frequency for Kozai-driven projectiles: Target bodies on inclined circular orbits}}

\author[1]{Youpeng Liang}
\author[1]{Junhai Huang}
\author[1,2]{Xiaodong Liu\thanks{E-mail: liuxd36@mail.sysu.edu.cn}}
\affil[1]{School of Aeronautics and Astronautics, Shenzhen Campus of Sun Yat-sen University, Shenzhen, Guangdong 518107, China}
\affil[2]{Shenzhen Key Laboratory of Intelligent Microsatellite Constellation, Shenzhen, Guangdong 518107, China}

\date{}
\begin{document}
\maketitle

\begin{abstract}
{\color{black}
Existing Öpik-type collision-frequency methods already incorporate the Kozai-driven secular evolution of the orbital elements of high-inclination projectiles.
However, these methods generally assume that the target body's orbit lies in the reference plane defined by the orbital plane of the perturbing body.
The present paper extends the semi-analytical framework developed by \citet{Vokrouhlicky2012} for a target body on a circular orbit in the reference plane to the case of a target body on a circular orbit with a non-zero inclination relative to that plane.
The target body's nodal precession rate is prescribed to be constant and may be zero.
In this geometry, whether the two orbits intersect depends not only on the secular state of the projectile's orbit but also on the relative nodal longitude between the two orbits.
To account for this dependence, the relative nodal longitude is introduced as an additional geometrical variable, and the framework is extended accordingly.
When the target body's circular orbit lies in the reference plane, the present formulation analytically reduces to the zero-inclination case described by \citet{Vokrouhlicky2012}.
The numerical results also confirm this reduction.
For the two cases with inclined target-body orbits, the collision frequencies computed with the present framework are used to predict semi-analytical decay curves for the fraction of projectiles remaining.
These predicted curves closely match those obtained from direct dynamical simulations.
}

\end{abstract}


\begin{figure*}
\centering
\includegraphics[width=1.0\textwidth,trim=1pt 1pt 1pt 1pt,clip]{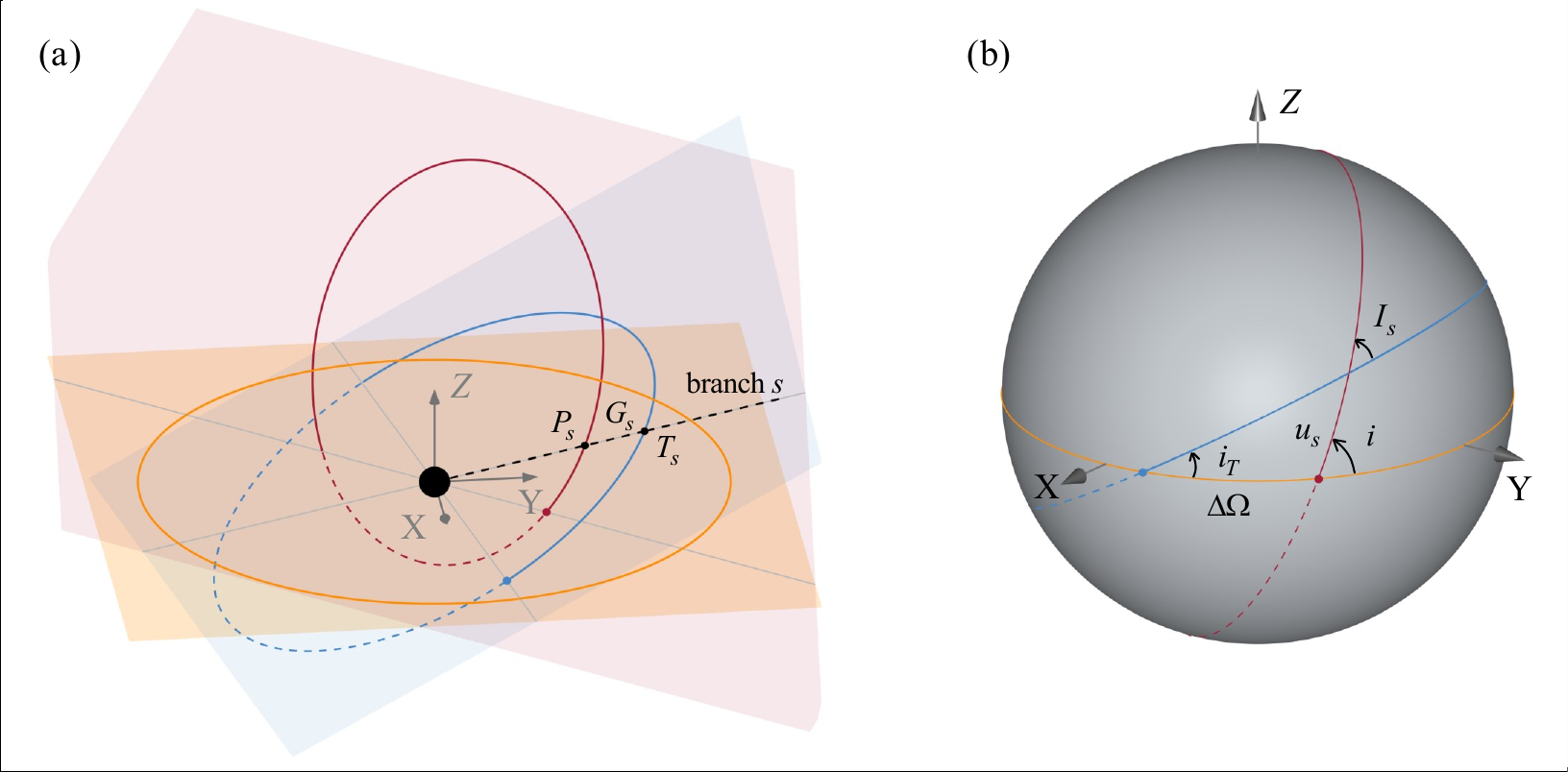}
\caption{
Schematic illustration of the orbital geometry and associated angular variables used in this work.
Panel~(a) shows the three-dimensional geometry of the orbits and their orbital planes.
The black sphere at the origin represents the central body.
The orange, red, and blue orbital curves, together with the translucent planes of the same colours, represent the orbits and orbital planes of the perturbing body, the projectile, and the target body, respectively.
The red and blue dots mark the ascending nodes of the projectile's orbit and the target body's orbit, respectively.
The perturbing body's orbital plane is adopted as the reference plane.
The black dashed segment indicates the direction along branch \(s\) of the mutual line of nodes.
Along this branch, the marked points \(P_s\) and \(T_s\) lie on the projectile's orbit and the target body's orbit, respectively.
Panel~(b) schematically illustrates the corresponding angular variables on the celestial sphere.
Following the same colour convention, the orange, red, and blue circles represent the projections of the three orbits on the celestial sphere.
On the celestial sphere, the red and blue dots indicate the directions of the ascending nodes of the projectile's orbit and the target body's orbit, respectively.
The angle \(u_s\) denotes the projectile's argument of latitude at \(P_s\) along the depicted mutual-node branch, measured from the ascending-node direction of the projectile's orbit.
The angle \(\Delta\Omega\) denotes the relative nodal longitude between the projectile's orbit and the target body's orbit.
The angles \(i\) and \(i_T\) denote the inclinations of the projectile's orbit and the target body's orbit relative to the reference plane, respectively.
The angle \(I_s\) denotes the mutual inclination associated with branch~\(s\).
}

\label{fig:system_geometry}
\end{figure*}

\section{Introduction}
{\color{black}}
Collision frequency is widely used in celestial mechanics and planetary science to study the collisional evolution of small-body populations, impact fluxes onto planets and satellites, and the transport and loss of dust and meteoroid populations.
Although direct dynamical simulations can in principle estimate collision frequencies from ensembles of projectiles, doing so reliably generally requires long integrations and repeated checks for close encounters or collisions between each projectile and the target body.

Öpik-type collision-frequency methods provide an efficient semi-analytical approach for estimating collision frequencies.
The original formulation by \citet{Opik1951} estimated the collision frequency for encounters between a target body on a circular orbit and projectiles moving on bound Keplerian orbits about the same central body.
Later, \citet{Wetherill1967} and \citet{Greenberg1982} extended this approach to cases in which both the target body and the projectile move on eccentric Keplerian orbits.
In these classical Öpik-type methods, the semimajor axes, eccentricities, and inclinations are treated as fixed parameters, whereas the longitude of ascending node and argument of periapsis are assumed to be uniformly distributed over \([0,2\pi)\).
Under these assumptions, the collision frequency is not obtained by tracking individual collision events directly.
Instead, it is evaluated by combining the geometrical conditions for close orbital encounters with the probability that the two bodies simultaneously occupy the corresponding locations on their respective orbits.

Under Kozai-driven secular evolution, a projectile's eccentricity and inclination can undergo coupled oscillations, while its longitude of ascending node and argument of periapsis generally evolve non-uniformly in time.
To account for this behaviour, \citet{Vokrouhlicky2012} generalized the Öpik-type framework by tracking the projectile's orbital elements over a Kozai cycle.
These time-dependent orbital elements specify the projectile's instantaneous orbital geometry throughout the cycle, thereby allowing the orbital-intersection configurations that contribute to the collision frequency to be identified.

The corresponding local collision contributions are then averaged over the Kozai cycle to obtain the long-term mean collision frequency.
\citet{Pokorny2013} subsequently extended this approach to a target body on an eccentric orbit undergoing prescribed apsidal precession.
In the formulations developed by \citet{Vokrouhlicky2012} and \citet{Pokorny2013}, the target body's orbit is confined to the reference plane defined by the orbital plane of the perturbing body.

Under this planar constraint, the projectile's orbit can intersect the target body's orbit only at the projectile's ascending or descending node with respect to the reference plane.
The orbital-intersection condition can therefore be determined from the projectile's secular orbital state alone.
By contrast, when the target body's orbit is inclined relative to the reference plane, the orbital-intersection condition also depends on the relative nodal longitude between the two orbits.

The present paper extends the Öpik-type framework developed by \citet{Vokrouhlicky2012} to a target body moving on an inclined circular orbit with a prescribed constant nodal precession rate that may be zero.
In this formulation, the relative nodal longitude between the two orbits enters the orbital-intersection condition as an additional geometrical variable.

The paper is organized as follows.
Section~2 develops the semi-analytical framework.
Section~3 derives its reduction when the target body's orbit lies in the reference plane and validates the reduced formulation against \citet{Vokrouhlicky2012}.
Section~4 applies the framework to two cases with inclined target-body orbits and compares the semi-analytical results with those from direct dynamical simulations.
Section~5 summarizes the conclusions.

\section{Semi-analytical collision-frequency framework}
{\color{black}
Following the treatment developed by \citet{Vokrouhlicky2012}, the collision-frequency calculation is decomposed into slow- and fast-variable components by exploiting the separation between the shorter Keplerian orbital timescales and the longer secular timescales.
The projectile's orbital elements that undergo Kozai-driven secular variation and the target body's longitude of ascending node, which evolves at the prescribed constant nodal-precession rate, are treated as slow variables. By contrast, the orbital phases of the two bodies along their respective instantaneous Keplerian orbits are treated as fast variables.
At the slow-variable level, the temporal weight of configurations in which the two instantaneous orbits satisfy the required proximity condition during the secular evolution is evaluated, thereby yielding the slow-variable probability factor.
At the fast-variable level, the probability that the target body's fast orbital phase lies within the range of phases allowed for collision when the projectile reaches the corresponding orbital location is evaluated, thereby yielding the fast-phase probability factor.
The product of these two probability factors gives the collision contribution associated with each relevant configuration.
The mean collision frequency is then obtained by averaging the accumulated contributions over multiple Kozai cycles and dividing the result by the projectile's Keplerian orbital period.

The orbital geometry of a physical system consisting of a central body, a perturbing body, a projectile, and a target body is illustrated schematically in panel~(a) of Fig.~\ref{fig:system_geometry}.
The perturbing body moves on a circular orbit about the central body, and its orbital plane defines the reference plane.
All inclinations used below are measured with respect to this plane.

The projectile is treated as a massless point particle with negligible physical radius, and its long-term orbital evolution is driven by the quadrupole-order, doubly averaged secular perturbation of the perturbing body.
The target body is modelled as a dynamically massless object moving on a prescribed inclined circular orbit about the central body.
Its semimajor axis and inclination are fixed, while its longitude of ascending node evolves in physical time at a prescribed constant nodal precession rate that may be zero.
The target body's geometrical radius is denoted by \(R_T\).
In the present framework, gravitational focusing is neglected, and the collision radius is therefore set equal to the geometrical radius, \(R=R_T\).
}

\subsection{Time-parametrized secular evolution of the projectile}

{\color{black}
After averaging over the mean anomalies of the projectile and the perturbing body and retaining the quadrupole-order term, the projectile's secular dynamics is governed by a normalized Kozai Hamiltonian.
In this approximation, the semimajor axis \(a\) of the projectile's orbit is conserved.
The dimensionless component of the projectile's specific orbital angular momentum along the normal to the reference plane is also conserved and is referred to as the Kozai constant:
\begin{equation}
c=\sqrt{1-e^2}\cos i ,
\end{equation}
where \(e\) and \(i\) are the eccentricity and inclination of the projectile's orbit, respectively.
The conservation of \(c\) constrains the coupled secular evolution of the eccentricity and inclination of the projectile's orbit.

The argument of periapsis \(\omega\) of the projectile's orbit is combined with its eccentricity \(e\) to define the nonsingular eccentricity variables:
\begin{equation}
k=e\cos\omega, \qquad h=e\sin\omega .
\end{equation}
The inverse relations are:
\begin{equation}
e=\sqrt{k^2+h^2}, \qquad
\omega=\operatorname{atan2}(h,k)\quad (e>0).
\label{eq:e_omega_from_kh}
\end{equation}
When \(e=0\), \(k=h=0\), and \(\omega\) is undefined.
For notational compactness, define:
\begin{equation}
g\equiv\sqrt{1-e^2}
=\sqrt{1-k^2-h^2}.
\label{eq:g_definition}
\end{equation}
With this notation, the Kozai constant defined above can be written as \(c=g\cos i\), from which the orbital inclination can be recovered as:
\begin{equation}
\cos i=\frac{c}{g}.
\label{eq:inclination_from_cg}
\end{equation}

The quadrupole-order, doubly averaged secular dynamics of the projectile are described by the classical Kozai--Lidov model \citep{Kozai1962,Lidov1962}, with the normalized Hamiltonian adopted from \citet{Vokrouhlicky2012}:
\begin{equation}
\begin{aligned}
H(k,h;c)
={}&
\left[2+3(k^2+h^2)\right]
\left(\frac{3c^2}{g^2}-1\right)
\\
&+15(k^2-h^2)
\left(1-\frac{c^2}{g^2}\right).
\end{aligned}
\label{eq:kozai_hamiltonian}
\end{equation}

The resulting secular evolution equations for the nonsingular variables \(k\) and \(h\), expressed in terms of the dimensionless secular time \(\tau\), are
\begin{equation}
\begin{aligned}
\frac{{\rm d}k}{{\rm d}\tau}
&=
-\frac{\partial H}{\partial h}
=
\frac{12h}{g^4}
\left[3g^4-5c^2(1-k^2)\right],
\\
\frac{{\rm d} h}{{\rm d}\tau}
&=
\frac{\partial H}{\partial k}
=
\frac{12k}{g^4}
\left(2g^4+5c^2h^2\right).
\end{aligned}
\label{eq:secular_evolution_kh}
\end{equation}

Because \(c\) is conserved and \(H(k,h;c)\) has no explicit dependence on \(\tau\), the total derivative of the Hamiltonian \(H\) along the secular evolution of \(k\) and \(h\) is
\begin{equation}
\frac{{\rm d} H}{{\rm d}\tau}
=
\frac{\partial H}{\partial k}\frac{{\rm d}k}{{\rm d}\tau}
+
\frac{\partial H}{\partial h}\frac{{\rm d} h}{{\rm d}\tau}
=0 .
\end{equation}
Consequently, for an initial state \((k_0,h_0)\), the pair \((k(\tau),h(\tau))\) remains on the level curve
\begin{equation}
H(k,h;c)=H(k_0,h_0;c).
\label{eq:hamiltonian_level_curve}
\end{equation}

The evolution equation for the projectile's longitude of ascending node \(\Omega\), derived by \citet{KinoshitaNakai2007} using the same normalized Hamiltonian as in the present formulation, is
\begin{equation}
\frac{{\rm d} \Omega}{{\rm d}\tau}
=
-\frac{1}{g}\frac{\partial H}{\partial c}
=
-\frac{12c}{g^3}\left(1-k^2+4h^2\right).
\label{eq:dOmega_dtau}
\end{equation}

The dimensionless secular time \(\tau\) is related to the physical time \(t\) by the time reparametrization given by \citet{Vokrouhlicky2012}:
\begin{equation}
{\rm d}t
=
\frac{16}{\gamma_\star g}\,{\rm d}\tau ,
\label{eq:time_conversion_dt_dtau}
\end{equation}
where \(\gamma_\star\) is the characteristic frequency scale associated with the quadrupole-order secular perturbation. For the circular orbit of the perturbing body considered here, \(\gamma_\star\) is
\begin{equation}
\gamma_\star
=
\frac{\mu_P}{a_P^3}
\sqrt{\frac{a^3}{\mu_0}} ,
\end{equation}
where \(\mu_0\) and \(\mu_P\) are the gravitational parameters of the central and perturbing bodies, respectively, and \(a_P\) is the semimajor axis of the perturbing body's orbit.

Integrating Eq.~\eqref{eq:time_conversion_dt_dtau} from a reference value \(\tau_0\) of the dimensionless secular time, with \(t_0\equiv t(\tau_0)\) denoting the corresponding reference physical time, gives
\begin{equation}
t(\tau)
=
t_0+
\int_{\tau_0}^{\tau}
\frac{16}{\gamma_\star g(\tau')}\,{\rm d}\tau' .
\label{eq:time_conversion_Integrating}
\end{equation}

The target body's longitude of ascending node is prescribed as a function of physical time \(t\):
\begin{equation}
\Omega_T(t)
=
\Omega_{T,0}
+
\dot{\Omega}_T(t-t_0),
\label{eq:OmegaT_prescribed_precession}
\end{equation}
where \(\dot{\Omega}_T\) is the prescribed constant nodal precession rate of the target body. It may be positive, negative, or zero; a negative value
corresponds to nodal regression.

Along the \(\tau\)-parametrized secular evolution, the relative nodal longitude between the projectile's orbit and the target body's orbit is
\begin{equation}
\Delta\Omega(\tau)
=
\Omega(\tau)
-
\Omega_T[t(\tau)] .
\label{eq:DeltaOmega_definition}
\end{equation}

{\color{black}
Taken together, Eqs.~\eqref{eq:secular_evolution_kh}, \eqref{eq:dOmega_dtau}, \eqref{eq:time_conversion_Integrating}, \eqref{eq:OmegaT_prescribed_precession}, and \eqref{eq:DeltaOmega_definition} determine a time-ordered curve
\(\bigl(k(\tau),h(\tau),\Delta\Omega(\tau)\bigr)\)
in the three-dimensional secular parameter space
\((k,h,\Delta\Omega)\).
Although the projectile's nodal evolution is determined by Eq.~\eqref{eq:dOmega_dtau}, the Hamiltonian level condition in Eq.~\eqref{eq:hamiltonian_level_curve} does not involve \(\Delta\Omega\).
Therefore, in the three-dimensional secular parameter space, this condition defines a cylindrical Hamiltonian level surface obtained by extending the closed Hamiltonian level curve in the \((k,h)\) plane along the \(\Delta\Omega\) direction.
As illustrated in Fig.~\ref{fig:extended_secular_parameter_space_it1}, the time-ordered curve lies on this surface, and its projection onto the \((k,h)\) plane coincides with the closed Hamiltonian level curve.
}

}

\subsection{Determination of exact orbital-intersection roots}
{\color{black}
As illustrated in panel~(a) of Fig.~\ref{fig:system_geometry}, the line of intersection of the projectile's orbital plane and the target body's orbital plane is referred to as the mutual line of nodes.
For non-coplanar orbital planes, this line is uniquely determined.
Its two opposite directions are referred to as the two mutual-node branches.
The branch directed along the cross product of the target body's orbital-plane unit normal and the projectile's orbital-plane unit normal, taken in that order, is labelled \(s=+1\), whereas the opposite branch is labelled \(s=-1\).
Any intersection point of two non-coplanar Keplerian orbits must lie on the mutual line of nodes.
The orbital-intersection problem can therefore be reduced to comparing the radial distances of the two orbits along each mutual-node branch.

Along mutual-node branch \(s\), let \(P_s\) and \(T_s\) denote the intersections of this branch with the projectile's orbit and the target body's orbit, respectively.
The radial distance of \(P_s\) from the central body is denoted by \(r_s\), whereas the radial distance of \(T_s\) is \(a_T\) because the target body's orbit is circular.
Let \(u_s\) denote the argument of latitude of the projectile at \(P_s\), measured from the ascending-node direction of the projectile's orbit, as illustrated in panel~(b) of Fig.~\ref{fig:system_geometry}.
The signed radial-distance difference along this branch is then
\begin{equation}
G_s
=
r_s-a_T
=
\frac{a g^2}
{1+k\cos u_s+h\sin u_s}
-
a_T,
\qquad s=\pm1.
\label{eq:Gs_radial_distance}
\end{equation}
Accordingly, the exact orbital-intersection condition along mutual-node branch \(s\) can be written simply as \(G_s=0\).

Expressing the direction of mutual-node branch \(s\) in the projectile's orbital plane gives the corresponding argument of latitude as
\begin{equation}
\begin{aligned}
u_s
&=
\operatorname{atan2}
\left[
s\sin i_T\sin\Delta\Omega,\right.
\\
&\qquad\left.
s\left(
\sin i\cos i_T
-
\cos i\sin i_T\cos\Delta\Omega
\right)
\right],
\qquad s=\pm1.
\end{aligned}
\label{eq:us_orbital_elements}
\end{equation}

With the fixed parameters left implicit, Eqs.~\eqref{eq:g_definition}, \eqref{eq:inclination_from_cg}, and \eqref{eq:us_orbital_elements} allow the signed radial-distance difference \(G_s\) to be written as
\begin{equation}
G_s=G_s(k,h,\Delta\Omega),
\qquad s=\pm1.
\label{eq:Gs_extended_variables}
\end{equation}
Accordingly, its value along the time-ordered curve \(\bigl(k(\tau),h(\tau),\Delta\Omega(\tau)\bigr)\) in the extended secular parameter space is
\begin{equation}
G_s(\tau)
\equiv
G_s\!\left(
k(\tau),h(\tau),\Delta\Omega(\tau)
\right),
\qquad s=\pm1.
\label{eq:Gs_time_ordered_curve}
\end{equation}
Over a specified interval in \(\tau\), the values of \(\tau\) at which \(G_s(\tau)\) vanishes are termed exact orbital-intersection roots and are ordered separately on each branch by increasing \(\tau\). The \(w\)-th root on branch \(s\) is denoted by \(\tau_{s,w}\) and satisfies
\[G_s(\tau_{s,w})=0.\]
At \(\tau_{s,w}\), the two instantaneous orbits form an exact orbital-intersection configuration along branch \(s\).}

\begin{figure}
\centering
\includegraphics[width=1.0\columnwidth]{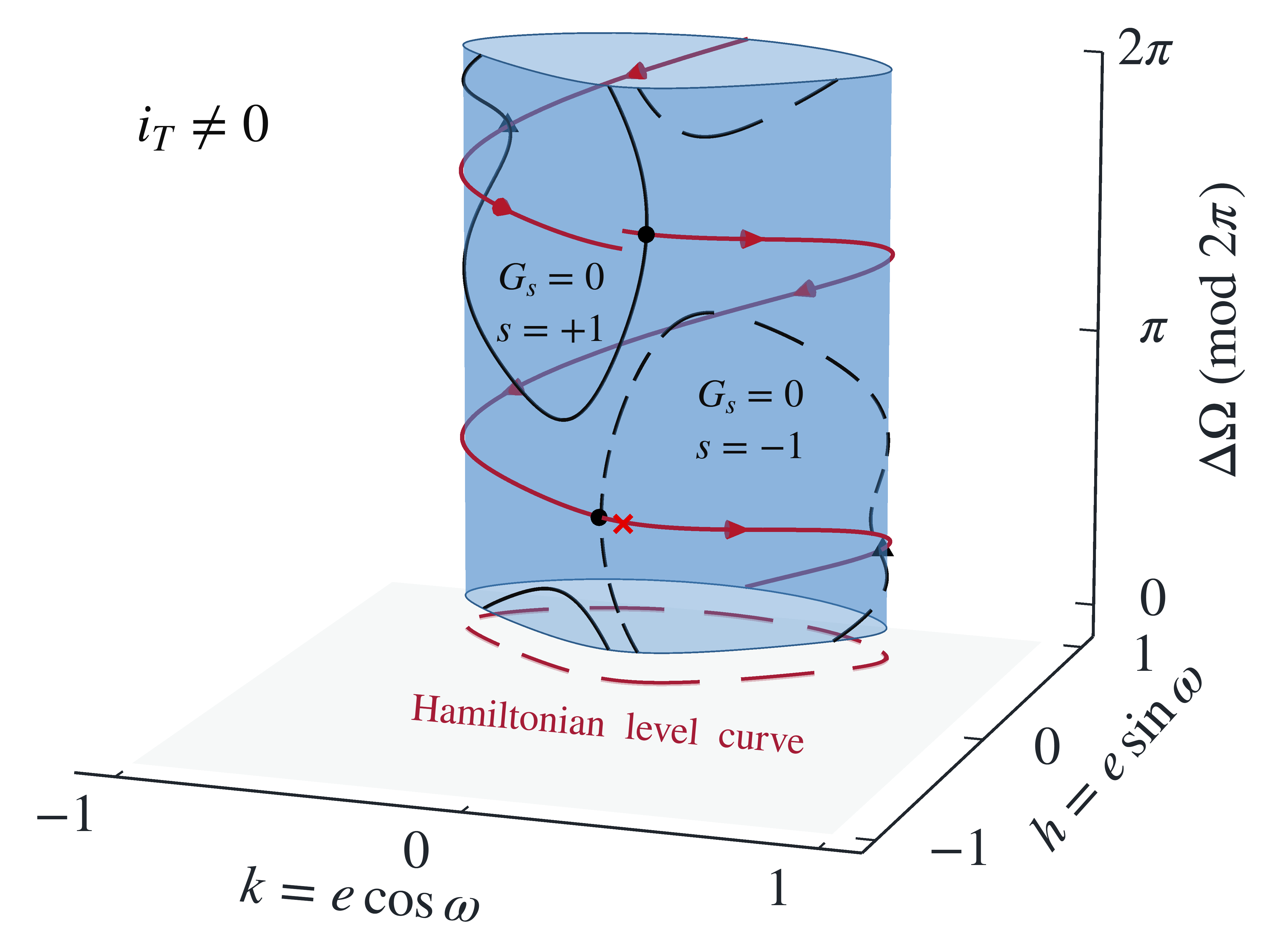}
\caption{{\color{black}
Time-ordered curve and exact orbital-intersection roots in the extended secular parameter space \((k,h,\Delta\Omega)\) for \(i_T\neq0\).
The red curve represents the time-ordered curve \(\bigl(k(\tau),h(\tau),\Delta\Omega(\tau)\bigr)\) over two consecutive complete Lidov--Kozai cycles.
The arrows indicate the direction of increasing \(\tau\), and the red cross marks the boundary between the two cycles.
The red dashed curve is the projection of the time-ordered curve onto the horizontal \((k,h)\) plane and coincides with the closed Hamiltonian level curve.
Extending this closed curve along the \(\Delta\Omega\) direction produces the translucent cylindrical Hamiltonian level surface.
The black solid and dashed curves on this surface are the zero contours \(G_s=0\) for the mutual-node branches \(s=+1\) and \(s=-1\), respectively.
For \(i_T\neq0\), these zero contours generally vary with \(\Delta\Omega\).
The black circles and triangles mark the intersections of these zero contours with the time-ordered curve during the first and second complete Lidov--Kozai cycles, respectively.
Each marker represents the secular state at an exact orbital-intersection root.
}}
\label{fig:extended_secular_parameter_space_it1}
\end{figure}

{\color{black}
Figure~\ref{fig:extended_secular_parameter_space_it1} provides a geometric interpretation of the exact orbital-intersection roots in the extended secular parameter space.
On the cylindrical Hamiltonian level surface introduced above, the conditions \(G_s=0\) define zero contours for the two mutual-node branches.
For \(i_T\neq0\), these zero contours generally depend on \(\Delta\Omega\) and therefore vary along the \(\Delta\Omega\) direction.
Their intersections with the time-ordered curve correspond to the secular states at exact orbital-intersection roots.
The zero contours and their intersections with the time-ordered curve are included in the figure solely to visualize this correspondence.
In the actual calculation, the exact orbital-intersection roots are obtained directly, for each branch, by locating the zeros of the one-dimensional function \(G_s(\tau)\) along the time-ordered curve; constructing the zero contours or searching for their intersections over the Hamiltonian level surface is not required.
}

\subsection{Probability associated with local time windows}

{\color{black}
For a non-zero collision radius \(R\), a local time window is defined around a root \(\tau_{s,w}\), within which the local spatial separation between the two instantaneous orbits near the corresponding mutual node remains smaller than \(R\).

Near the mutual node associated with branch \(s\), the two instantaneous Keplerian orbital arcs are approximated by their local tangent lines; 
\(D_s\) denotes the minimum separation between these lines. 
In the notation adopted here, the local relation given by \citet{Greenberg1982} connects \(D_s\) to the signed radial-distance difference \(G_s\) through the dimensionless geometrical projection factor \(B_s\):
\begin{equation}
\label{eq:local_minimum_distance_greenberg}
D_s = B_s\,|G_s| .
\end{equation}
This relation provides a local straight-line estimate of the separation near the mutual node, rather than the exact minimum separation between the corresponding Keplerian orbital arcs.

Specializing the local relation of \citet{Greenberg1982} to the orbital geometry considered here yields the following expression for the projection factor \(B_s\):
\begin{equation}
\label{eq:geometrical_projection_factor}
B_s
=
\frac{
\left|\sin I_s\right|
}{
\left[
\sin^2 I_s
+
\left(
\frac{
k\sin u_s-h\cos u_s
}{
1+k\cos u_s+h\sin u_s
}
\right)^2
\right]^{1/2}
}.
\end{equation}
Here \(I_s\) denotes the mutual inclination associated with branch \(s\), as illustrated in panel~(b) of Fig.~\ref{fig:system_geometry}. Its magnitude is determined by
\begin{equation}
\cos I_s
=
\cos i\cos i_T
+
\sin i\sin i_T\cos\Delta\Omega .
\label{eq:I_define}
\end{equation}
The sign of \(I_s\) is defined according to the crossing direction of the projectile's instantaneous Keplerian orbit at the mutual node associated with branch \(s\): \(I_s>0\) for an upward crossing of the target body's orbital plane and \(I_s<0\) for a downward crossing.

Following the local time-window treatment of \citet{Vokrouhlicky2012}, the signed radial-distance difference \(G_s(\tau)\) is expanded in \(\tau\) about the root \(\tau_{s,w}\). Since \(G_s(\tau_{s,w})=0\), retaining the first-order term gives
\begin{equation}
G_s(\tau)
\simeq
\left.
\frac{{\rm d}G_s}{{\rm d}\tau}
\right|_{\tau=\tau_{s,w}}
\left(\tau-\tau_{s,w}\right).
\label{eq:Gs_linearized_near_root}
\end{equation}
Here \({\rm d}G_s/{\rm d}\tau\) is the total derivative along the time-ordered curve:
\begin{equation}
\frac{{\rm d}G_s}{{\rm d}\tau}
=
\frac{\partial G_s}{\partial k}
\frac{{\rm d}k}{{\rm d}\tau}
+
\frac{\partial G_s}{\partial h}
\frac{{\rm d}h}{{\rm d}\tau}
+
\frac{\partial G_s}{\partial\Delta\Omega}
\frac{{\rm d}\Delta\Omega}{{\rm d}\tau}.
\end{equation}
In this expression, the derivatives \({\rm d}k/{\rm d}\tau\) and \({\rm d}h/{\rm d}\tau\) are given by Eq.~\eqref{eq:secular_evolution_kh}. 
For the relative nodal derivative \(\frac{{\rm d}\Delta\Omega}{{\rm d}\tau}\), differentiating Eqs.~\eqref{eq:DeltaOmega_definition} and \eqref{eq:OmegaT_prescribed_precession} with respect to \(\tau\) and using the time conversion in Eq.~\eqref{eq:time_conversion_dt_dtau} gives
\begin{equation}
\frac{{\rm d}\Delta\Omega}{{\rm d}\tau}
=
\frac{{\rm d}\Omega}{{\rm d}\tau}
-
\dot{\Omega}_T
\frac{{\rm d}t}{{\rm d}\tau}.
\label{eq:dDeltaOmega_dtau}
\end{equation}
The projectile nodal derivative \({\rm d}\Omega/{\rm d}\tau\) on the right-hand side is given by Eq.~\eqref{eq:dOmega_dtau}.

Within a local neighbourhood of the root \(\tau_{s,w}\), \(B_s(\tau)\) is approximated by its value at the root,
\begin{equation}
B_s(\tau)\simeq B_{s,w},
\qquad
B_{s,w}\equiv B_s(\tau_{s,w}).
\end{equation}
Together with the first-order expansion of \(G_s(\tau)\) in Eq.~\eqref{eq:Gs_linearized_near_root}, this approximation gives the local separation near \(\tau_{s,w}\) as
\begin{equation}
\begin{aligned}
D_s(\tau)
&=
B_s(\tau)\left|G_s(\tau)\right|
\\
&\simeq
B_{s,w}
\left|
\left.
\frac{{\rm d}G_s}{{\rm d}\tau}
\right|_{\tau=\tau_{s,w}}
\right|
\left|\tau-\tau_{s,w}\right|.
\end{aligned}
\label{eq:Ds_linearized_near_root}
\end{equation}

The local time window is defined by the condition
\begin{equation}
D_s(\tau)\leq R.
\label{eq:collision_radius_requirement}
\end{equation}
Substitution of Eq.~\eqref{eq:Ds_linearized_near_root} into this condition gives the linearized half-width in normalized secular time:
\begin{equation}
\Delta\tau_{s,w}^{\rm lin}
=
\frac{
R
}{
B_{s,w}
\left|
\left.
\frac{{\rm d}G_s}{{\rm d}\tau}
\right|_{\tau=\tau_{s,w}}
\right|
}.
\label{eq:linearized_time_window_half_width}
\end{equation}
The corresponding candidate boundaries are
\begin{equation}
\tau_{s,w}^{{\rm lin},\pm}
=
\tau_{s,w}
\pm
\Delta\tau_{s,w}^{\rm lin}.
\label{eq:linearized_time_window_boundaries}
\end{equation}

The local approximations, however, do not necessarily reproduce the variation of \(D_s(\tau)\) near \(\tau_{s,w}\) with sufficient accuracy.
The consistency of the two candidate boundaries with the defining condition \(D_s(\tau)=R\) is therefore assessed before the candidate window is adopted.

For this purpose, the relative residuals at the two candidate boundaries are written compactly as
\begin{equation}
\epsilon_{s,w}^{\pm}
=
\frac{
B_s\!\left(\tau_{s,w}^{{\rm lin},\pm}\right)
\left|
G_s\!\left(\tau_{s,w}^{{\rm lin},\pm}\right)
\right|
-R
}{
R
}.
\label{eq:linearized_window_boundary_residuals}
\end{equation}
The maximum relative boundary residual is then defined as
\begin{equation}
\epsilon_{s,w}^{\max}
=
\max\!\left(
\left|\epsilon_{s,w}^{-}\right|,
\left|\epsilon_{s,w}^{+}\right|
\right).
\label{eq:linearized_window_boundary_deviation}
\end{equation}

For a prescribed tolerance \(\epsilon_{\rm tol}\), the linearized candidate boundaries are accepted when \(\epsilon_{s,w}^{\max}\leq\epsilon_{\rm tol}\).
When \(\epsilon_{s,w}^{\max}>\epsilon_{\rm tol}\), they are replaced by the nearest solutions on either side of \(\tau_{s,w}\) of the boundary equation
\(B_s(\tau)\left|G_s(\tau)\right|=R\) located by an adaptive search.
The resulting boundaries are denoted by \(\tau_{s,w}^{-}\) and \(\tau_{s,w}^{+}\), and define the local time window
\begin{equation}
\tau
\in
\left[
\tau_{s,w}^{-},
\tau_{s,w}^{+}
\right].
\label{eq:final_local_time_window}
\end{equation}
The corresponding physical duration is obtained from the time conversion in Eq.~\eqref{eq:time_conversion_dt_dtau}:
\begin{equation}
\Delta t_{s,w}
=
\int_{\tau_{s,w}^{-}}^{\tau_{s,w}^{+}}
\frac{{\rm d}t}{{\rm d}\tau}
\,{\rm d}\tau .
\label{eq:physical_local_window_duration}
\end{equation}

The period of the Lidov--Kozai cycle in normalized secular time, denoted by \(\mathcal P_\tau\), is defined as the smallest positive return time of the secular state \((k,h)\) along the Hamiltonian level curve:
\begin{equation}
\left(
k(\tau_0+\mathcal P_\tau),
h(\tau_0+\mathcal P_\tau)
\right)
=
\left(
k(\tau_0),
h(\tau_0)
\right).
\label{eq:Kozai_period_tau_return}
\end{equation}

The corresponding physical period of the Lidov--Kozai cycle is also obtained from the time conversion in Eq.~\eqref{eq:time_conversion_dt_dtau}:
\begin{equation}
T_{\rm Kozai}
=
\int_{\tau_0}^{\tau_0+\mathcal P_\tau}
\frac{{\rm d}t}{{\rm d}\tau}
\,{\rm d}\tau .
\label{eq:physical_Kozai_period}
\end{equation}

The slow-variable probability factor associated with the local time window around \(\tau_{s,w}\) is defined by
\begin{equation}
P_{1,s,w}
=
\frac{\Delta t_{s,w}}{T_{\rm Kozai}} .
\label{eq:P1_time_window_probability}
\end{equation}
}

\subsection{Probability associated with the target body's fast phase}
{\color{black}
Geometrical proximity between the two instantaneous orbits is necessary for collision, but does not by itself ensure that the two bodies collide.
When the projectile passes through the vicinity of the mutual node associated with \(\tau_{s,w}\), a collision further requires the simultaneous presence of the target body in this vicinity, thereby restricting its fast orbital phase to an allowed interval.
Assuming that this phase is uniformly distributed along the target body's circular orbit, the fraction of a complete orbital-phase cycle occupied by this interval defines the fast-phase probability factor \(P_{2,s,w}\).

In the treatment of \citet{Greenberg1982}, the two bodies are approximated as moving with constant velocities along the local tangents to their instantaneous Keplerian orbits near the mutual node.
Specializing the resulting probability expression to the orbital geometry considered here and expressing it in the present notation gives
\begin{equation}
P_{2,s,w}
=
\frac{R}{4a_T}
\frac{U_{s,w}}{U_{\perp,s,w}} .
\label{eq:P2_inclined_target_vector}
\end{equation}
Here \(U_{s,w}\) is the magnitude of the relative velocity between the projectile and the target body at the mutual node associated with \(\tau_{s,w}\), and \(U_{\perp,s,w}\) is the magnitude of the component of this relative velocity perpendicular to the local tangent to the target body's orbit.

Using the exact orbital-intersection condition and the Keplerian velocity relations, Eq.~\eqref{eq:P2_inclined_target_vector} can be written entirely in scalar form as
\begin{equation}
P_{2,s,w}
=
\frac{R}{4a_T}
\left(
\frac{3-T_{s,w}}{2-F_{s,w}}
\right)^{1/2}.
\label{eq:P2_inclined_target_scalar}
\end{equation}
The dimensionless auxiliary quantities \(T_{s,w}\) and \(F_{s,w}\) are defined by
\begin{equation}
\begin{aligned}
T_{s,w}
&=
\frac{a_T}{a}
+
2
\left(
\frac{a g_{s,w}^2}{a_T}
\right)^{1/2}
\cos I_{s,w},
\\
F_{s,w}
&=
\frac{a_T}{a}
+
\frac{a g_{s,w}^2}{a_T}
\cos^2 I_{s,w}.
\end{aligned}
\label{eq:Tsw_Fsw_scalar}
\end{equation}
Here \(g_{s,w}\) and \(I_{s,w}\) denote \(g\) and \(I_s\), respectively, evaluated at the exact orbital-intersection root \(\tau_{s,w}\).
}

\subsection{Mean collision frequency over multiple cycles}
{\color{black}
During one complete Lidov--Kozai cycle, the secular variables \(k(\tau)\) and \(h(\tau)\) return to their initial values on the Hamiltonian level curve, whereas the relative nodal longitude \(\Delta\Omega(\tau)\) does not necessarily return to its initial value modulo \(2\pi\).
Let \(\delta_\Omega\) denote the increment in the relative nodal longitude accumulated over one complete Lidov--Kozai cycle:
\begin{equation}
\delta_\Omega
\equiv
\Delta\Omega(\tau_0+\mathcal P_\tau)
-
\Delta\Omega(\tau_0).
\label{eq:relative_nodal_advance_per_cycle}
\end{equation}
Accordingly, the three variables \(\bigl(k(\tau),h(\tau),\Delta\Omega(\tau)\bigr)\) return simultaneously to their initial values after \(N\) complete Lidov--Kozai cycles if and only if the positive integer \(N\) satisfies
\begin{equation}
N\delta_\Omega
\equiv
0
\pmod{2\pi}.
\label{eq:simultaneous_return_condition}
\end{equation}

Exact commensurability between \(\delta_\Omega\) and \(2\pi\) requires an exact relation among the parameters of the specified system and thus constitutes a special case.
The discussion therefore begins with the incommensurate case, \(\delta_\Omega/(2\pi)\notin\mathbb Q\).
In this case, the relative nodal longitude evaluated at the same phase of successive Lidov--Kozai cycles takes the values
\[
\Delta\Omega\!\left(\tau_0+n\mathcal P_\tau\right)
\bmod 2\pi,
\qquad
n=0,1,2,\ldots,
\]
which do not repeat and are equidistributed over \([0,2\pi)\).
Because the exact orbital-intersection roots \(\tau_{s,w}\) and the associated probability factors \(P_{1,s,w}\) and \(P_{2,s,w}\) depend on \(\Delta\Omega\), a calculation restricted to a single Lidov--Kozai cycle does not account for the variation in orbital-intersection configurations from one cycle to the next.
Accordingly, the collision frequency must be evaluated over multiple complete Lidov--Kozai cycles.

For \(N\) consecutive complete Lidov--Kozai cycles beginning at \(\tau_0\), let \(\mathcal W^{(N)}\) denote the index set of all exact orbital-intersection roots contained within these cycles:
\begin{equation}
\mathcal W^{(N)}
=
\left\{
(s,w):
s=\pm1,\,
\tau_{s,w}\in
\left[
\tau_0,\,
\tau_0+N\mathcal P_\tau
\right)
\right\}.
\label{eq:root_index_set_multicycle}
\end{equation}

For each root \(\tau_{s,w}\) with \((s,w)\in\mathcal W^{(N)}\), the corresponding dimensionless root-level contribution is defined as
\begin{equation}
\Pi_{s,w}
=
P_{1,s,w}P_{2,s,w}.
\label{eq:root_contribution}
\end{equation}

The cumulative mean collision frequency over these \(N\) complete Lidov--Kozai cycles, denoted by \(\Gamma^{(N)}\), is defined as
\begin{equation}
\Gamma^{(N)}
=
\frac{1}{N T_{\rm orb}}
\sum_{(s,w)\in\mathcal W^{(N)}}
P_{1,s,w}P_{2,s,w}.
\label{eq:multi_period_collision_frequency}
\end{equation}
Here, \(T_{\rm orb}=2\pi\sqrt{a^3/\mu_0}\) is the projectile's Keplerian orbital period. The factor \(1/N\) averages the accumulated dimensionless root-level contributions over the \(N\) complete Lidov--Kozai cycles, while multiplication by \(T_{\rm orb}^{-1}\) converts the resulting dimensionless average into a collision frequency.

The cumulative intrinsic collision probability over these \(N\) complete Lidov--Kozai cycles, denoted by \(p^{(N)}\), is defined as
\begin{equation}
p^{(N)}
=
\frac{\Gamma^{(N)}}{R^2}.
\label{eq:intrinsic_collision_probability_N}
\end{equation}

As \(N\) increases, the convergence of \(\Gamma^{(N)}\), or equivalently of \(p^{(N)}\), is assessed.
If these sequences converge, their limiting values define the mean collision frequency and the intrinsic collision probability, respectively:
\begin{equation}
\Gamma=\lim_{N\to\infty}\Gamma^{(N)},\qquad
p=\lim_{N\to\infty}p^{(N)}
=\lim_{N\to\infty}\frac{\Gamma^{(N)}}{R^2}.
\label{eq:collision_frequency_and_intrinsic_probability_limits}
\end{equation}

In numerical calculations, once further increases in \(N\) produce negligible changes, the values of \(\Gamma^{(N)}\) and \(p^{(N)}\) at a sufficiently large finite \(N\) are adopted as approximations to \(\Gamma\) and \(p\), respectively.

In the commensurate case, let \(N_{\rm rep}\) denote the smallest positive integer satisfying Eq.~\eqref{eq:simultaneous_return_condition}.
After \(N_{\rm rep}\) complete Lidov--Kozai cycles, the three variables \(\bigl(k(\tau),h(\tau),\Delta\Omega(\tau)\bigr)\) return simultaneously to their initial values, and their evolution is periodic with period \(N_{\rm rep}\mathcal P_\tau\).
The exact orbital-intersection roots occurring during these \(N_{\rm rep}\) cycles, together with their associated root-level contributions, therefore constitute one complete repeating sequence.
The mean collision frequency and the intrinsic collision probability are then given by the cumulative quantities evaluated over these \(N_{\rm rep}\) cycles:
\begin{equation}
\Gamma
=
\Gamma^{(N_{\rm rep})},
\qquad
p
=
p^{(N_{\rm rep})}.
\label{eq:commensurate_collision_frequency}
\end{equation}}

\section{Reduction to the \texorpdfstring{$i_T=0$}{iT=0} case}

\subsection{Analytical reduction}
{\color{black}
This subsection examines how the semi-analytical framework developed in Section~2 reduces when the target body's orbital inclination is set to \(i_T=0\).
This condition is assumed throughout, and the resulting formulation is compared with the corresponding formulation developed by \citet{Vokrouhlicky2012}, with differences in notation and symbol conventions taken into account.

First, the projectile's secular dynamics---including the Hamiltonian in Eq.~\eqref{eq:kozai_hamiltonian}, the evolution of \(k(\tau)\) and \(h(\tau)\) in Eq.~\eqref{eq:secular_evolution_kh}, the projectile's nodal precession rate in Eq.~\eqref{eq:dOmega_dtau}, and the conversion between normalized secular time and physical time in Eq.~\eqref{eq:time_conversion_dt_dtau}---remain unchanged, because none of these relations involves the target body's orbital elements.

Second, with the target body's orbital plane coincident with the reference plane, the mutual line of nodes between the two orbital planes reduces to the nodal line of the projectile's orbit with respect to the reference plane.
The two mutual-node branches point toward the projectile's ascending and descending nodes, respectively.

Applying the expression for the projectile's argument of latitude at the mutual nodes, Eq.~\eqref{eq:us_orbital_elements}, to this geometry gives \(u_+=0\) and \(u_-=\pi\). Therefore,
\begin{equation}
\cos u_s=s,
\qquad
\sin u_s=0,
\qquad
s=\pm1.
\label{eq:us_trig_iT_zero}
\end{equation}

Substituting Eq.~\eqref{eq:us_trig_iT_zero} into the signed radial-distance difference \(G_s\) defined in Eq.~\eqref{eq:Gs_radial_distance} gives
\begin{equation}
\left.G_s\right|_{i_T=0}
=
\frac{a g^2}{1+s k}
-
a_T,
\qquad
s=\pm1.
\label{eq:Gs_iT_zero}
\end{equation}
{\color{black}
This equation is independent of \(\Delta\Omega\), so, for any fixed value of \(\Delta\Omega\), the condition \(\left.G_s\right|_{i_T=0}=0\) is satisfied at the same \((k,h)\) coordinates.
Figure~\ref{fig:extended_secular_parameter_space_it0} illustrates the resulting geometry: on the cylindrical Hamiltonian level surface, the corresponding zero contours are vertical lines parallel to the \(\Delta\Omega\) axis.
}

\begin{figure}
\centering
\includegraphics[width=1.0\columnwidth]{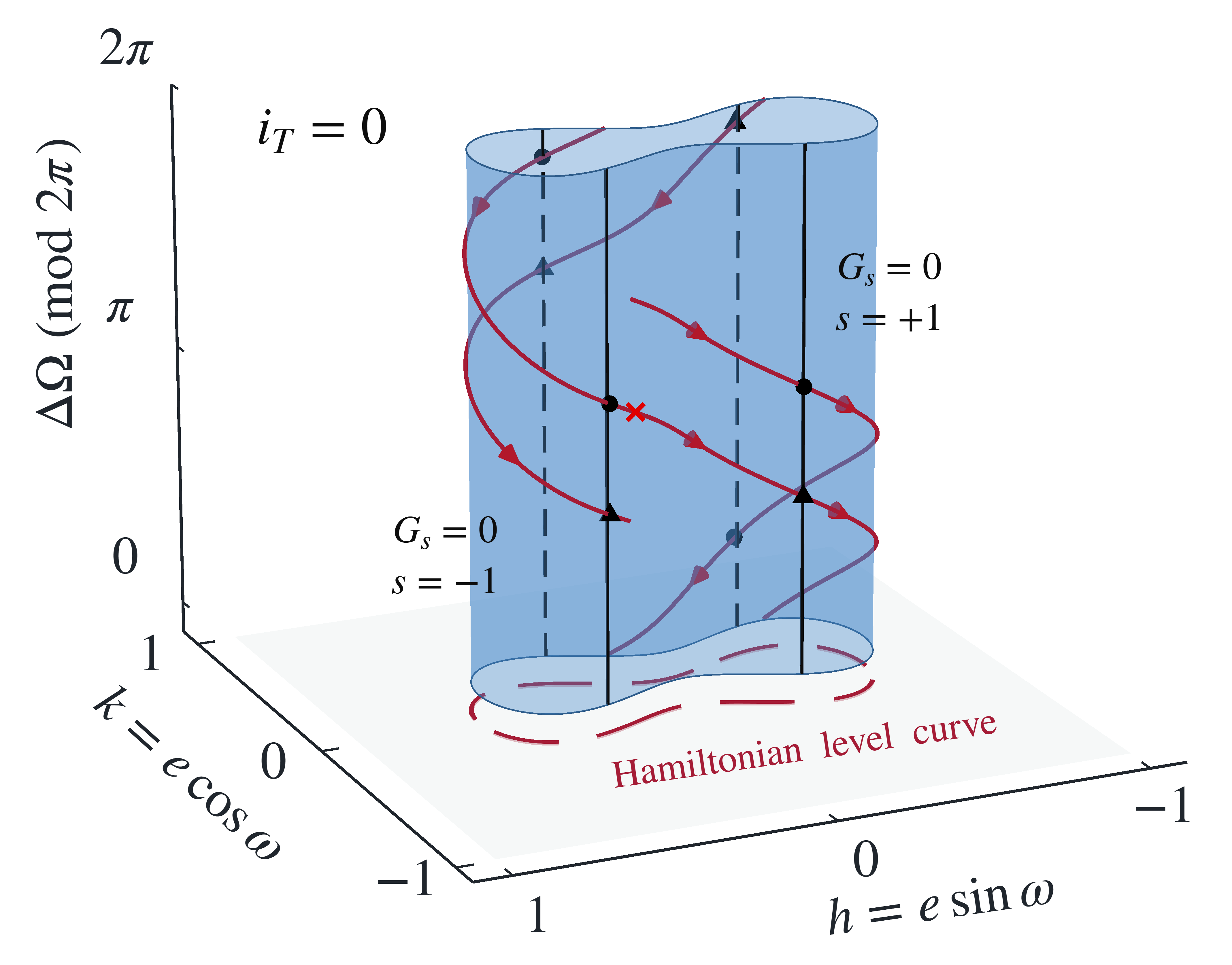}
\caption{{\color{black}
Time-ordered curve and exact orbital-intersection roots in the extended secular parameter space \((k,h,\Delta\Omega)\) for \(i_T=0\).
The geometrical construction and graphical conventions are the same as in Fig.~\ref{fig:extended_secular_parameter_space_it1}.
The projectile initial conditions differ from those in Fig.~\ref{fig:extended_secular_parameter_space_it1}, so the Hamiltonian level curves and corresponding cylindrical level surfaces also differ.
}}
\label{fig:extended_secular_parameter_space_it0}
\end{figure}

Third, the linearized half-width in Eq.~\eqref{eq:linearized_time_window_half_width} depends on the geometrical projection factor \(B_s\) and the derivative \({\rm d}G_s/{\rm d}\tau\) at each exact orbital-intersection root \(\tau_{s,w}\).

The geometrical projection factor \(B_s\), given by Eq.~\eqref{eq:geometrical_projection_factor}, involves the mutual inclination \(I_s\) and the projectile's argument of latitude \(u_s\) at the corresponding mutual node.
Setting \(i_T=0\) in the mutual-inclination relation, Eq.~\eqref{eq:I_define}, gives
\begin{equation}
\left.\cos I_s\right|_{i_T=0}
=
\cos i.
\label{eq:I_cos_iT_zero}
\end{equation}
Since \(0\leq i\leq\pi\), it follows that
\begin{equation}
\left.\lvert\sin I_s\rvert\right|_{i_T=0}
=
\sin i.
\label{eq:I_sin_iT_zero}
\end{equation}
Only \(\lvert\sin I_s\rvert\) and \(\sin^2 I_s\) enter the expression for \(B_s\), so its value is unaffected by the branch-dependent sign of \(I_s\).

Substituting Eqs.~\eqref{eq:I_sin_iT_zero} and \eqref{eq:us_trig_iT_zero} into the expression for the geometrical projection factor \(B_s\) in Eq.~\eqref{eq:geometrical_projection_factor} gives
\begin{equation}
\left.B_s\right|_{i_T=0}
=
\frac{(1+s k)\sin i}
{\left[
h^2+(1+s k)^2\sin^2 i
\right]^{1/2}},
\qquad
s=\pm1.
\label{eq:Bs_iT_zero}
\end{equation}

The derivative required in Eq.~\eqref{eq:linearized_time_window_half_width} follows by differentiating Eq.~\eqref{eq:Gs_iT_zero} along the evolution of \(k(\tau)\) and \(h(\tau)\).
Together with Eq.~\eqref{eq:Bs_iT_zero}, it determines the linearized local time window around each root \(\tau_{s,w}\).
The corresponding physical duration \(\Delta t_{s,w}\) and slow-variable probability factor \(P_{1,s,w}\) remain given by Eqs.~\eqref{eq:physical_local_window_duration} and \eqref{eq:P1_time_window_probability}, respectively.

Fourth, for \(i_T=0\), substituting Eq.~\eqref{eq:I_cos_iT_zero} into Eq.~\eqref{eq:Tsw_Fsw_scalar} gives the following reduced forms of the auxiliary quantities \(T_{s,w}\) and \(F_{s,w}\) entering the fast-phase probability factor \(P_{2,s,w}\):
\begin{equation}
\begin{aligned}
\left.T_{s,w}\right|_{i_T=0}
&=
\frac{a_T}{a}
+
2
\left(
\frac{a g_{s,w}^2}{a_T}
\right)^{1/2}
\cos i,
\\
\left.F_{s,w}\right|_{i_T=0}
&=
\frac{a_T}{a}
+
\frac{a g_{s,w}^2}{a_T}
\cos^2 i.
\end{aligned}
\label{eq:Tsw_Fsw_iT_zero}
\end{equation}
These reduced expressions directly yield the corresponding reduced form of \(P_{2,s,w}\).

{\color{black}
Finally, because \(k(\tau)\) and \(h(\tau)\) repeat over each complete Lidov--Kozai cycle and \(G_s\) is independent of \(\Delta\Omega\), the exact orbital-intersection roots recur at the same relative phases within successive cycles.
The black circles and triangles in Fig.~\ref{fig:extended_secular_parameter_space_it0} illustrate this recurrence in the first and second cycles, respectively; corresponding roots have the same \((k,h)\) coordinates, although their values of \(\Delta\Omega\) may differ.
The preceding reductions further imply that the corresponding probability factors \(P_{1,s,w}\) and \(P_{2,s,w}\), and hence the root-level contributions \(\Pi_{s,w}=P_{1,s,w}P_{2,s,w}\), are identical in every cycle.
}

Analogously to the cumulative root index set \(\mathcal W^{(N)}\) defined in Eq.~\eqref{eq:root_index_set_multicycle}, let \(\mathcal W_{(N)}\) denote the index set of all exact orbital-intersection roots contained in the \(N\)-th complete Lidov--Kozai cycle:
\begin{equation}
\mathcal W_{(N)}
=
\left\{
(s,w):
s=\pm1,\,
\tau_{s,w}\in
\left[
\tau_0+(N-1)\mathcal P_\tau,\,
\tau_0+N\mathcal P_\tau
\right)
\right\}.
\label{eq:root_index_set_individual_cycle}
\end{equation}
Here \(\mathcal W^{(N)}\) indexes the roots occurring during the first \(N\) complete cycles, whereas \(\mathcal W_{(N)}\) indexes those occurring during the \(N\)-th cycle alone.

The individual-cycle collision frequency \(\Gamma_{(N)}\) and the corresponding individual-cycle intrinsic collision probability \(p_{(N)}\) are defined as
\begin{equation}
\Gamma_{(N)}
=
\frac{1}{T_{\rm orb}}
\sum_{(s,w)\in\mathcal W_{(N)}}
\Pi_{s,w},
\qquad
p_{(N)}
=
\frac{\Gamma_{(N)}}{R^2}.
\label{eq:individual_cycle_collision_quantities}
\end{equation}
The parenthesized subscript identifies an individual cycle, whereas the previously defined \(\Gamma^{(N)}\) and \(p^{(N)}\), given by Eqs.~\eqref{eq:multi_period_collision_frequency} and \eqref{eq:intrinsic_collision_probability_N}, respectively, denote the cumulative results over the first \(N\) complete cycles.
Equivalently,
\begin{equation}
\Gamma^{(N)}
=
\frac{1}{N}
\sum_{n=1}^{N}
\Gamma_{(n)},
\qquad
p^{(N)}
=
\frac{1}{N}
\sum_{n=1}^{N}
p_{(n)}.
\label{eq:cumulative_from_individual_cycles}
\end{equation}

In the \(i_T=0\) case, the cycle-to-cycle repetition yields the following relations for every positive integer \(N\) and every \(n\) satisfying \(1\leq n\leq N\):
\begin{equation}
\Gamma^{(N)}
=
\Gamma_{(n)}
=
\Gamma,
\qquad
p^{(N)}
=p_{(n)}
=
p.
\label{eq:cycle_quantities_iT_zero}
\end{equation}
Thus, all individual-cycle quantities are identical and equal to both the corresponding cumulative quantities and their limiting values.

The preceding analysis derives the \(i_T=0\) forms of the individual components of the present semi-analytical framework.
Indeed, the reduced formulation recovers the mathematical construction and resulting expressions given by \citet{Vokrouhlicky2012}, with differences confined to notation and symbol conventions.}

\subsection{Numerical validation}
{\color{black}
The two cases considered by \citet{Vokrouhlicky2012} are used to validate the present framework.
In both cases, the central, perturbing, and target bodies are the Sun, Jupiter, and Earth, respectively, with \(a_P=5.2\,{\rm AU}\) and \(a_T=1\,{\rm AU}\).
The dynamically massless target body follows a prescribed circular orbit in the reference plane, with its radius set to \(R=4.26\times10^{-4}\,{\rm AU}\); gravitational focusing is neglected. 
The projectile has \(a=1.4\,{\rm AU}\), \(e_0=0.2\), and \(i_0=65^\circ\).
The projectile's initial argument of periapsis is set to \(\omega_0=0^\circ\) and \(60^\circ\) in Cases 1 and 2, respectively.
To maintain direct correspondence with the formulation developed by \citet{Vokrouhlicky2012}, the local time-window boundaries in these two validation cases are taken directly from the linearized estimates, without applying the adaptive search.

\begingroup
\makeatletter
\def\fps@table{b}
\makeatother
\begin{table}
\centering
\caption{Comparison of the cumulative intrinsic collision probabilities \(p^{(100)}\) obtained with the present framework and the intrinsic collision probabilities \(p\) reported by \citet{Vokrouhlicky2012}. The values are given in units of \({\rm AU^{-2}\,yr^{-1}}\).}
\label{tab:vokrouhlicky2012_validation}
\footnotesize
\renewcommand{\arraystretch}{1.15}
\renewcommand{\tabularxcolumn}[1]{m{#1}}
\begin{tabularx}{1.0\columnwidth}
{@{}*{4}{>{\centering\arraybackslash}X}@{}}
\hline
Case
& \(p\)
& \(p^{(100)}\)
& \shortstack{Relative\\difference} \\
\hline
Case 1
& \(0.586\)
& \(0.58754\)
& \(0.26\%\) \\
Case 2
& \(0.413\)
& \(0.41338\)
& \(0.09\%\) \\
\hline
\end{tabularx}
\end{table}
\endgroup

For each case, the calculation is performed over 100 complete Lidov--Kozai cycles, yielding the cumulative intrinsic collision probability \(p^{(100)}\).
The resulting values are compared in Table~\ref{tab:vokrouhlicky2012_validation} with the corresponding intrinsic collision probabilities reported by \citet{Vokrouhlicky2012}.
The relative differences are \(0.26\%\) and \(0.09\%\) for Cases 1 and 2, respectively, indicating close numerical agreement.

The cycle-to-cycle equality established analytically in Eq.~\eqref{eq:cycle_quantities_iT_zero} is also examined numerically using \(p_{(N)}\) and \(p^{(N)}\).
Fig.~\ref{fig:iT0_cycle_consistency} shows, for both validation cases, the signed relative deviations \((p_{(N)}-p^{(100)})/p^{(100)}\) and \((p^{(N)}-p^{(100)})/p^{(100)}\).

\begin{figure}
\centering
\includegraphics[width=\columnwidth,trim=0 15pt 0 0,clip]{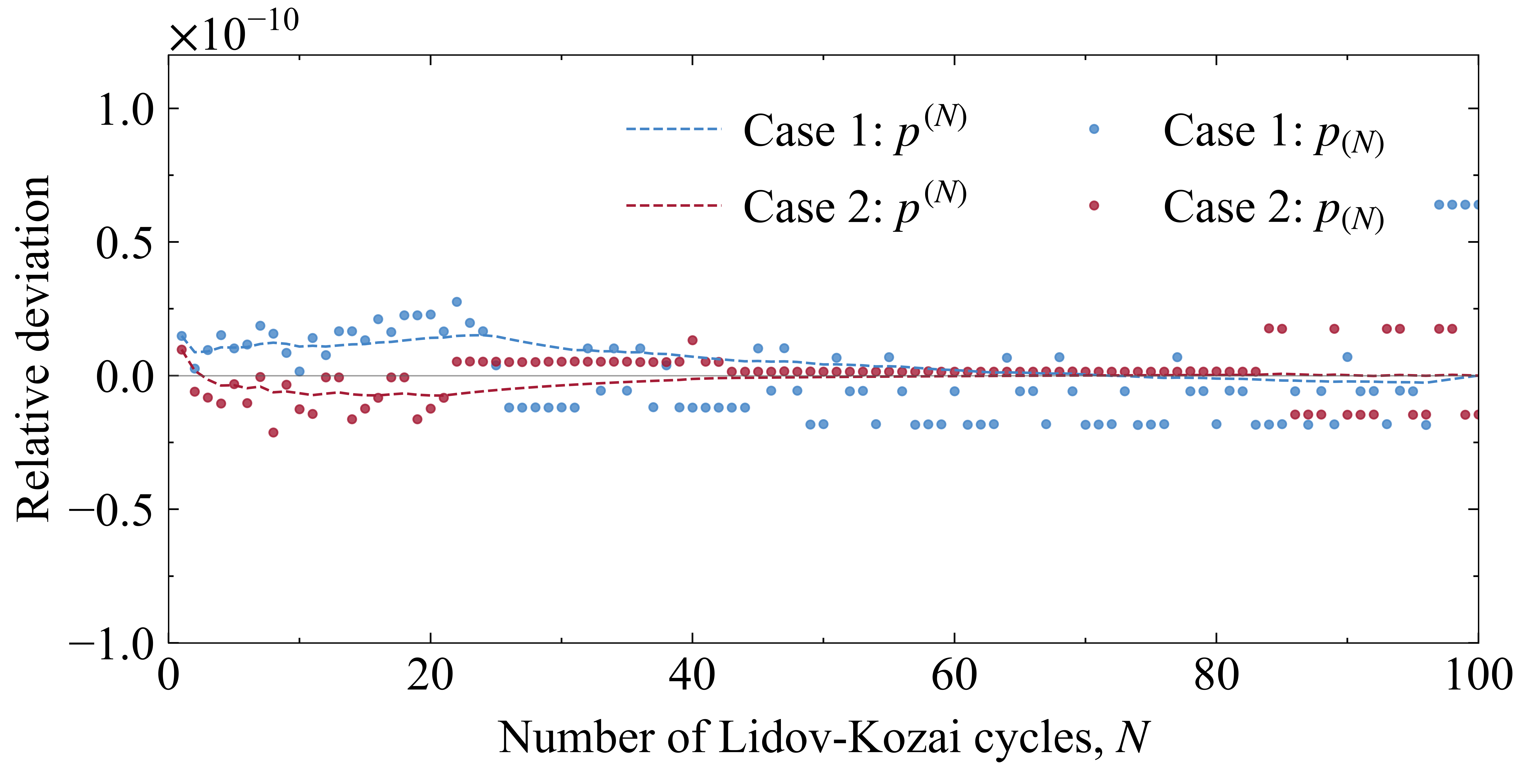}
\caption{
Cycle-to-cycle consistency for the two validation cases with \(i_T=0\).
The dashed curves and circular markers show the signed relative deviations of the cumulative intrinsic collision probability \(p^{(N)}\) and the individual-cycle intrinsic collision probability \(p_{(N)}\), respectively, from the corresponding \(p^{(100)}\) value.
The deviations are calculated using the unrounded \(p^{(100)}\) values.
}
\label{fig:iT0_cycle_consistency}
\end{figure}

For both validation cases, the magnitudes of the signed relative deviations of \(p_{(N)}\) and \(p^{(N)}\) remain at approximately \(10^{-10}\) or below.
The cycle-to-cycle equality in Eq.~\eqref{eq:cycle_quantities_iT_zero} is therefore recovered to numerical precision.}

\section{Inclined-target results and dynamical validation}

\subsection{Validation cases}
{\color{black}
Two inclined-target cases are considered to compare the multi-cycle semi-analytical results with direct REBOUND integrations.

In both cases, the central and perturbing bodies are the Sun and Jupiter, respectively, with \(a_P=5.2\,{\rm AU}\).
The target body follows a prescribed circular orbit with \(a_T=1\,{\rm AU}\) and \(\Omega_{T,0}=0^\circ\).
The projectile has \(a=1.4\,{\rm AU}\), \(e_0=0.2\), \(i_0=65^\circ\), and \(\Omega_0=0^\circ\).
With gravitational focusing neglected, the collision radius is set equal to the target-body radius, \(R=R_T=8.527\times10^{-4}\,{\rm AU}\).
This enlarged value is adopted to improve the collision statistics in the direct integrations and is also used in the semi-analytical calculation.
The tolerance governing the adaptive search for the local time-window boundaries is set to \(\epsilon_{\rm tol}=0.03\) in both cases.

Case~3 adopts \(i_T=10^\circ\), \(\omega_0=20^\circ\), and \(\dot{\Omega}_T=0\).
Case~4 adopts \(i_T=30^\circ\), \(\omega_0=135^\circ\), and a uniform nodal regression rate \(\dot{\Omega}_T=-{2\pi}/{150000\,{\rm yr}}\).

\subsection{Multi-cycle semi-analytical results}

For each case, the cumulative mean collision frequency \(\Gamma^{(N)}\) defined in Eq.~\eqref{eq:multi_period_collision_frequency} is evaluated for \(N\leq500\).
Fig.~\ref{fig:inclined_prefix_convergence} shows the resulting sequences of \(\Gamma^{(N)}\).

In both cases, \(\Gamma^{(N)}\) stabilizes as \(N\) increases.
At \(N=500\), the values of \(\Gamma^{(500)}\) are \(3.80701\times10^{-7}\,{\rm yr}^{-1}\) and \(2.21227\times10^{-7}\,{\rm yr}^{-1}\) for Cases~3 and 4, respectively.
These values are adopted as numerical approximations to the corresponding mean collision frequencies \(\Gamma\).

\begin{figure}
\centering
\includegraphics[width=\columnwidth,trim=0 15pt 0 0,clip]{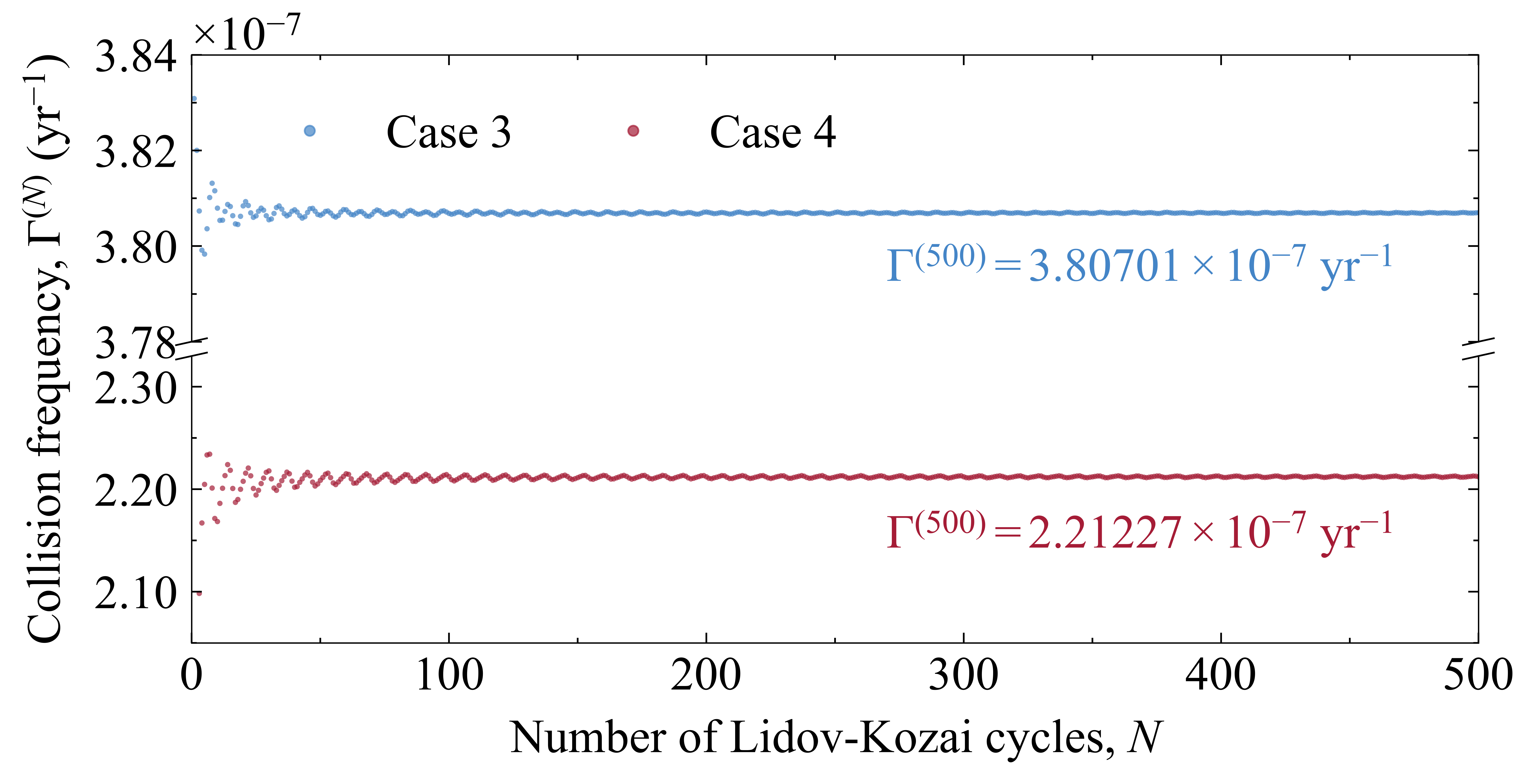}
\caption{
Cumulative mean collision frequency \(\Gamma^{(N)}\) as a function of the number \(N\) of complete Lidov--Kozai cycles for Cases~3 and 4.
The blue and red series correspond to Cases~3 and 4, respectively.
A broken vertical axis is used to display the two sequences at their respective scales.
}
\label{fig:inclined_prefix_convergence}
\end{figure}
}

\subsection{Direct dynamical validation}
{\color{black}
The direct integrations are performed with REBOUND \citep{ReinLiu2012} using the orbital parameters and collision radius specified above.
The Sun and Jupiter are included as massive gravitating bodies, whereas the projectiles are treated as massless test particles.
The target body is also treated as dynamically massless.
Its position is prescribed as a function of time according to the circular orbit and nodal evolution specified for each case, rather than determined by the REBOUND integration.

For each case, the direct integrations include \(N_0=1000\) projectiles.
All projectiles are assigned the same initial values of \(a\), \(e\), \(i\), \(\Omega\), and \(\omega\) as those adopted in the corresponding semi-analytical calculation. Each projectile is assigned an initial mean anomaly \(M\) sampled uniformly over \([0,2\pi)\).
The WHFast symplectic integrator \citep{ReinTamayo2015} is used with a fixed timestep of \(2\,{\rm days}\), and the integrations are carried out for \(10^7\,{\rm yr}\).
A collision with the target body is recorded when the projectile--target centre-to-centre separation becomes smaller than \(R_T\).
Collisions are detected using REBOUND's \texttt{linetree} collision-search algorithm, and the corresponding projectile is removed once a collision is identified.

Let \(N_{\rm rem}(t)\) denote the number of projectiles remaining at time \(t\).
The fraction of projectiles remaining is then
\begin{equation}
S(t)
=
\frac{N_{\rm rem}(t)}{N_0}.
\label{eq:projectile_remaining_fraction}
\end{equation}

For a constant mean collision frequency \(\Gamma\), the time evolution of the theoretical fraction of projectiles remaining, denoted by \(S_{\rm th}(t)\), is governed by
\begin{equation}
\frac{{\rm d}S_{\rm th}}{{\rm d}t}
=
-\Gamma S_{\rm th},
\qquad
S_{\rm th}(0)=1.
\label{eq:theoretical_remaining_fraction_evolution}
\end{equation}
The resulting time dependence is
\begin{equation}
S_{\rm th}(t)
=
\exp(-\Gamma t).
\label{eq:theoretical_remaining_fraction}
\end{equation}

Fig.~\ref{fig:inclined_remaining_fraction} shows the fraction of projectiles remaining obtained from the direct REBOUND integrations, together with the semi-analytical decay curve \(S_{\rm th}(t)\).
For both cases, the numerical fractions closely follow the corresponding semi-analytical decay curves.

\begin{figure}
\centering
\includegraphics[width=\columnwidth,trim=0 5pt 0 0,clip]{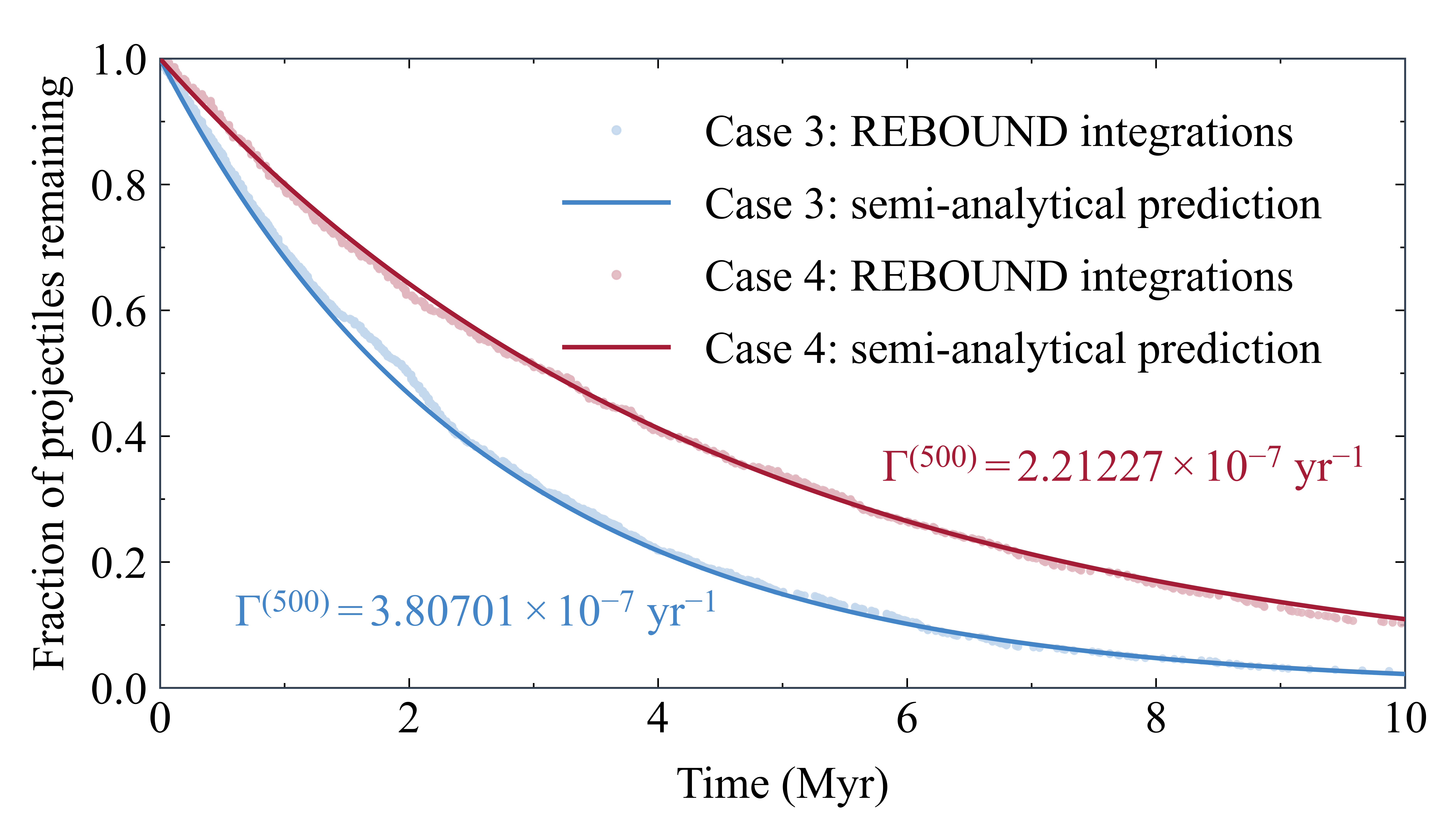}
\caption{
Fraction of projectiles remaining as a function of time for Cases~3 and 4.
The light-blue and light-red markers show the fractions obtained from the direct REBOUND integrations for Cases~3 and 4, respectively, whereas the blue and red solid curves show the corresponding semi-analytical predictions given by Eq.~\eqref{eq:theoretical_remaining_fraction}.}
\label{fig:inclined_remaining_fraction}
\end{figure}
}

\section{Conclusions}

{\color{black}

The present study extends the Öpik-type collision-frequency framework developed by \citet{Vokrouhlicky2012} for Kozai-driven, high-inclination projectiles to a target body moving on an inclined circular orbit. 
By treating the relative nodal longitude as an additional geometrical variable, the formulation accounts for the dependence of orbital intersection on the mutual orientation of the two orbital planes. 
The target body's orbital plane may remain fixed or undergo prescribed constant nodal precession.

When \(i_T=0\), the present semi-analytical framework reduces analytically to that of \citet{Vokrouhlicky2012} after differences in notation and symbol conventions are accounted for. 
This reduction is further verified through numerical comparisons.

For the two \(i_T\neq0\) cases examined, the cumulative mean collision frequency \(\Gamma^{(N)}\) stabilizes over the range of Lidov--Kozai cycles considered. 
Using these stabilized frequencies, the semi-analytical exponential-decay curves for the fraction of projectiles remaining closely follow the corresponding fractions obtained from the direct REBOUND integrations.
These results support the use of the present framework for estimating long-term collision frequencies for Kozai-driven projectiles encountering target bodies on prescribed inclined circular orbits.
}

\section*{Acknowledgements}
{\raggedright
This work was supported by the National Natural Science Foundation of China (Nos.~12472048, 12002397 and 62388101), and the Shenzhen Science and Technology Program (Grant No.~ZDSYS20210623091808026).
\par}

\section*{Data Availability}
The basic data of this work will be shared on reasonable request to the corresponding author.

\end{document}